\documentclass{article}

\usepackage[numbers]{natbib}         
\usepackage[colorlinks]{hyperref}    
\usepackage[english]{babel} 
\usepackage{amssymb}
\usepackage{amsmath}
\usepackage{txfonts}
\usepackage{mathdots}
\usepackage[classicReIm]{kpfonts}
\usepackage[dvips]{graphicx} 
\usepackage[a4paper, portrait, margin=1in]{geometry}
\usepackage{tabularx}
\usepackage{longtable}
\usepackage{multirow}
\usepackage{booktabs}
\usepackage[labelsep=period]{caption}
\usepackage{makecell}
\begin{document}
	\setlength{\parindent}{0pt}
	\setlength{\parskip}{1ex}
	
	\textbf{\Large Generative Adversarial Network for Image Synthesis}
	
	\bigbreak

	Yang Lei, Richard L.J. Qiu, Tonghe Wang, Walter J. Curran, Tian Liu and Xiaofeng Yang*
	
	Department of Radiation Oncology and Winship Cancer Institute, Emory University, Atlanta, GA 30322

	\bigbreak
	\bigbreak
	\bigbreak

	\textbf{*Corresponding author: }
	
	Xiaofeng Yang, PhD
	
	Department of Radiation Oncology
	
	Emory University School of Medicine
	
	1365 Clifton Road NE
	
	Atlanta, GA 30322
	
	E-mail: xiaofeng.yang@emory.edu

	\bigbreak
	\bigbreak
	\bigbreak
	\bigbreak
	\bigbreak
	\bigbreak

	\textbf{Abstract}

	This chapter reviews recent developments of generative adversarial networks (GAN)-based methods for medical and biomedical image synthesis tasks. These methods are classified into conditional GAN and Cycle-GAN according to the network architecture designs. For each category, a literature survey is given, which covers discussions of the network architecture designs, highlights important contributions and identifies specific challenges. 
	
	\bigbreak
	\bigbreak
	
	\textbf{keywords:} Image synthesis, deep learning, Generative Adversarial Network, GAN.

	\noindent 
	\section{ INTRODUCTION}
	
	Image synthesis is the process that generates synthetic/pseudo images in the target image modality/domain (named as target domain) from the inputs of source images that reside in a different image modality/domain (named as source domain). The aim of image synthesis is to bypass a certain imaging procedure and use the synthetic images instead. The motivation could be multifold: the specific image acquisition is infeasible; it bears additional labor and cost; some imaging procedures add ionizing radiation exposure to patients; uncertainties could be introduced from the image registration between different modalities. In recent years, research in image synthesis gains great interest in radiation oncology, radiology and biology \cite{RN75}. The presumed benefit has intrigued several investigations in a number of potential clinical applications such as magnetic resonance imaging (MRI)-only radiation therapy treatment planning \cite{RN52, RN22, RN10, RN19, RN14, RN62}, positron emission tomography (PET)/MRI scanning \cite{RN39, RN50}, proton stopping power estimation \cite{RN95, RN92, RN87, RN110}, synthetic image-aided auto-segmentation \cite{RN111, RN4, RN93, RN104, RN3, RN11, RN86}, low dose computerized tomography (CT) denoising \cite{RN59, RN13, RN3228}, image quality enhancement \cite{RN88, RN58, RN33, RN68}, reconstruction \cite{RN87, RN2}, high resolution visualization \cite{RN57} and etc. 
	
	Historically, image synthesis methods have been investigated for decades. The conventional methods, such as random forest-based method, dictionary learning-based method, etc., usually rely on models with explicit manually designed principles about the conversion of images from one modality to the other \cite{RN51, RN46,RN47, RN50, RN112, RN49, RN2006, RN48, RN52, RN1436}. Therefore, the conventional methods are usually application-specific and can be complicated \cite{RN89}.
	
	Unlike conventional machine learning, deep learning does not rely on hand-crafted features given by human \cite{RN101, RN84, RN89, RN75}. It utilizes neural network (NN) or convolutional NN (CNN) with several hidden layers containing a large number of neurons or convolutional kernels to automatically learn the way of extracting informative features. As a result, deep learning has been widely adopted in medical imaging and biomedical imaging field in the past several years \cite{RN105}. For image synthesis tasks, the workflow of deep learning-based methods usually consists of two stages: a training stage for the network to build the mapping between the source and target image domain; an inference stage to generate synthetic image, called as synthetic target image, from a new arrival source image. Various networks and architectures have been proposed for better performance on different tasks. In this literature survey, a class of network architectures, called as the generative adversarial networks (GANs), especially the conditional GAN (cGAN) \cite{RN20} and cycle consistent GAN (Cycle-GAN), are introduced and explained. The emerging GAN-based methods and applications geared for medical and biomedical image synthesis are systematically reviewed and discussed. In short, we aim to:
	
	•	Summarize the latest network architecture designs of cGAN and Cycle-GAN.
	
	•	Summarize the latest medical and biomedical image synthesis applications of cGAN and Cycle-GAN.
	
	•	Highlight important contributions and identify existing challenges.

	\noindent 
	\section{Literature Searching}

	The scope of this review is confined to both cGAN and Cycle-GAN methods designed for medical and biomedical image synthesis tasks. Medical image synthesis applications include studies about multi-modality MRI synthesis, proton stopping power estimation, image quality improvement/enhancement, super/high resolution visualization, MRI-only radiation therapy treatment planning, inter-modality image registration, segmentation, PET attenuation correction, and data augmentation for image analysis. Biomedical image synthesis applications include cell synthesis and tissue synthesis.
	
	Peer-reviewed journal and conference/proceeding publications were searched on PubMed using the criteria in title or abstract as of December 2020: (“pseudo” OR “synth*” OR “reconstruct*” OR “transform” OR “restor*” OR “correct*” OR “generat*”) AND “deep” AND “learning” AND “generative” OR “adversarial” OR “discriminat*” OR “cycle” OR “consistent” (“CT” OR “MR” OR “MRI” OR “PET” OR “Medical” OR “Biomedical” etc.).
	
	\noindent 
	\section{Network Architecture Design}
	
	The GAN, introduced by Ian J. Goodfellow et al. \cite{RN15}, is a developed approach of “generative modeling” using a flexible unsupervised deep learning architecture. Generative modeling uses unsupervised learning to automatically recognize and learn the patterns in the input data. The trained model can subsequently produce new data (output) that mimics the input data. Its ability of creating massive realistic contents makes it extremely popular and useful, gaining tremendous success in the field of computer vision. Naturally, the latest breakthrough of its supervised manners has been integrated into medical image synthesis as well.
	
	\noindent 
	\subsection{Conditional GAN}
	Traditional GAN was trained with no restrictions on data generation. Later, it was updated by using conditional image constraints to derive synthetic images with desired properties, coined as cGAN. CGANs have been extensively used in medical image synthesis studies due to their capability of data generation without explicitly modeling the probability density function. The adversarial loss brought by the discriminator provides a clever way of incorporating unlabeled samples into training and imposing higher order of consistency.
	
	CGAN is composed of a generative network and a discriminative network. The generative network is trained to generate synthetic images, and the discriminative network is trained to judge whether an input image is real or synthesized. The training goal of cGAN is to train the generative network to produce synthetic images that are realistic enough to fool the discriminator, and train the discriminative network to distinguish the synthetic images from real images. As the two networks play a zero-sum game, the performance of each one increases when they compete against each other until both networks reach their maximum potential. This conflict goal explains the name of “adversarial”. After the model is trained, the synthetic image of a new arrival source image can be obtained via feeding the source image into the trained generator network.

	\begin{figure}
		\centering
		\noindent \includegraphics*[width=6.50in, height=4.20in, keepaspectratio=true]{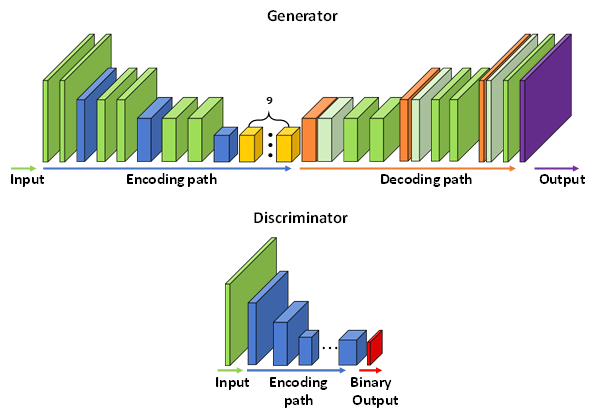}
		
		\noindent Fig. 1. An illustration of cGAN network architecture.
	\end{figure}
	
	Figure 1 shows an example of a cGAN. Basically, the generator network of cGAN can be implemented by an end-to-end fully convolutional network (FCN), such as U-Net-like architecture used in Figure 1, or be implemented by a non-end-to-end FCN. The end-to-end FCN can generate output that shares the same size as the input. The non-end-to-end FCN, on the other hand, can generate different sized output. The end-to-end FCN is often composed of an encoding path and a decoding path, where the encoding path down-samples the feature map size and the decoding path up-samples the feature map size to perform an end-to-end output. The encoding path is composed of several convolution layers with stride size of two or several convolution layers and followed by a max-pooling layer to reduce the feature maps’ size. The decoding path is composed of three deconvolution layers to obtain the end-to-end mapping, several convolution layers and a tanh layer to perform the regression. There may be several residual blocks \cite{RN21} or dense blocks \cite{RN23} used as short skip connection between the encoding path and the decoding path. Residual blocks are frequently used to learn the residual information between source and target image domain. Dense blocks are used to capture multi-scale or multi-level image features. For some applications, a long residual block is used as a long skip connection \cite{RN37}, which bypasses the feature maps from the first convolution layer to the last convolution layer, to guide all the hidden layers of the generator focusing on learning the difference between input source and target domain images. Some works integrated attention gates into the long skip connection of generator architecture to capture the most relevant semantic contextual information without enlarging the receptive field \cite{RN31}. The feature maps extracted from the coarse scale were used in gating to disambiguate irrelevant and noisy responses in long skip connections. This was performed immediately prior to the concatenation operation to merge only relevant activations. Additionally, attention gates filter the neuron activations during both the forward pass and the backward pass. The non-end-to-end FCN is commonly composed of an encoding path and maybe followed by several fully connect layers for the prediction task. The discriminator is often composed of several convolutional layers and max-pooling layers and followed by a sigmoid or soft-max layer to perform the binary classification.
	
	Many different variants of the cGAN framework were proposed to meet the desired output. In this study, we exams some cGAN frameworks that are or can be used for medical or biomedical image synthesis, which includes deep convolutional GAN (DCGAN), pix2pix and InfoGAN.

	\noindent 
	\subsubsection{DCGAN}
	
	DCGAN produces better and more stable training results when a fully connected layer is replaced by a fully convolutional layer. The architecture of the generator in DCGAN is illustrated in the work of \cite{RN16}. In the core of the framework, pooling layers were replaced with fractional-stride convolutions, which allowed it to learn from random input noise vector by own spatial upsampling to generate an image from it. There were two important changes adopted to modify the architecture of early cGAN, which were batch normalization and leaky ReLU. Batch normalization \cite{RN17} was used for regulating the poor initialization to prevent the deep generator from mode collapse which was a major drawback in the early GAN framework. Leaky ReLU \cite{RN19} activation was introduced at the place of maxout activation \cite{RN15} all layers of a discriminator which improved the resolution of image output.
	
	\noindent 
	\subsubsection{Pix2pix}
	
	The pix2pix is a supervised image-to-image translation model proposed by Isola et al.\cite{RN20}. It has received a multi-domain user acceptance in the computer vision community for image synthesis, whose merit is to combine the loss of cGAN with $l_1$-norm minimization loss (or termed as mean absolute error (MAE) loss) so that network learns not only the mapping from the input image to output image but also the loss function to generate the image resemble to the ground truth. To train this network, both adversarial loss of judging the authenticity of synthetic images and the image-based accuracy loss (such as MAE) are used. By using image-based accuracy loss, the cGAN is trained under a supervised manner, which can be more suitable for the image synthesis task when the learning targets are given. For example, in the image synthesis task of MRI-only radiation therapy, the paired planning CT and corresponding registered MRI are given for training. By training a supervised cGAN, the synthetic CT (sCT) for a new arrival MRI can not only look like a real CT but also has accurate intensity value, i.e., Hounsfield value (HU), which is essential for radiation therapy dose calculation.
	
	\noindent 
	\subsubsection{InfoGAN}
	For some medical image synthesis tasks, such as cone beam CT (CBCT) scatter correction and PET attenuation correction, the histogram of generated synthetic image also matters. If only using image intensity value accuracy as loss function, the model then cannot be supervised properly. Recently, InfoGAN was developed for computer vision tasks by adding an information-theoretic extension to the cGAN that is able to learn disentangled representations. InfoGAN is a cGAN that also maximizes the mutual information, which force the image distribution similarity between a small subset of the latent variables and the observation \cite{RN21}. A lower bound of the mutual information objective was derived that could be optimized efficiently. For example, InfoGAN successfully disentangled writing styles from digit shapes on the modified national institute of standards and technology (MNIST) dataset, pose from lighting of 3D rendered images, and background digits from the central digit on the street view house numbers (SVHN) dataset \cite{RN21}. It also discovered visual concepts that included hair styles, presence/absence of eyeglasses, and emotions on the CelebFaces attributes dataset. It would be expected that InfoGAN could serve well for medical or biomedical image synthesis in the future. Recently, InfoGAN was used for generating 3D CT volume from 2D projection data \cite{RN76}.

	\noindent 
	\subsection{Cycle-GAN}
	
	Many different variants of the Cycle-GAN framework were proposed for image synthesis. In this study, we review some Cycle-GAN frameworks that are or can be used for medical or biomedical image synthesis, which includes residual Cycle-GAN (Res-Cycle-GAN) \cite{RN21}, dense Cycle-GAN (Dense-Cycle-GAN) \cite{RN23}, UNIT \cite{RN22}, Bicycle-GAN \cite{RN23} and StarGAN \cite{RN24}.
	
	As introduced previously, cGANs, rely on two sub-networks, a generator and a discriminator that compete against each other, are optimized sequentially in a zero-sum framework. Cycle-GAN doubles the process of a typical cGAN by enforcing an inverse transformation, i.e. translating a synthetic target image back to source image domain, called as cycle source image \cite{RN103}. It further constrains the model and can increase the accuracy in output synthetic target image. In reality, mismatches could exist between source and target image domain for training set even after good image registration, which would cause ill-posed problem. To address this issue, Cycle-GAN introduces an additional cycle loop to force the model to be close to a one-to-one mapping.
	
	\begin{figure}
		\centering
		\noindent \includegraphics*[width=6.50in, height=4.20in, keepaspectratio=true]{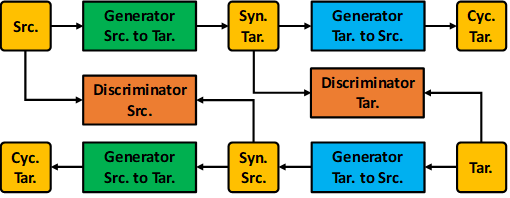}
		
		\noindent Fig. 2. An illustration of Cycle-GAN framework. Src. denotes the source image. Tar. denotes the target image. Syn. denotes the synthetic image. Cyc. denotes the cycle image.
	\end{figure}

	Figure 2 shows an example of the framework of traditional Cycle-GAN. As can be seen, the Cycle-GAN is composed of two full loop: the first loop is a mapping from source image domain to target image domain and then a mapping from target image domain back to source image domain; the second loop is a mapping from target image domain to source image domain and then a mapping from source image domain back to target image domain. Thus, the Cycle-GAN is composed of two generators, i.e., from source image domain to target image domain and from target image domain to source image domain. The two generators often share same network architecture, but with different parameters that are optimized alternately and independently. It also includes two discriminators: one is to judge whether the synthetic source image is real or fake; the other one is used to judge whether the synthetic target image is real or fake.
	
	\noindent 
	\subsubsection{Res-Cycle-GAN}
	
	For image synthesis tasks, promising results were accomplished by Cycle-GAN with residual blocks when source and target image modalities shared good similarity, such as  CBCT and CT images \cite{RN21}, low dose PET and full dose PET \cite{RN8}, and etc. Several residual blocks were used as short skip connections in generators of Cycle-GAN. Each residual block was constructed with a residual connection and multiple hidden layers, as shown in Figure 3. The input feature map extracted from source image bypassed the hidden layers of a residual block via the residual connection, therefore the hidden layers were assigned to learn the differences between source image and target image. A residual block was engineered using two convolution layers within residual connection and an element-wise sum operator.
	
	\begin{figure}
		\centering
		\noindent \includegraphics*[width=4in, height=3in, keepaspectratio=true]{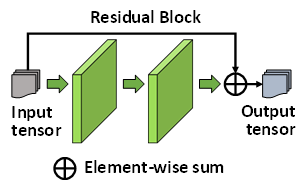}
		
		\noindent Fig. 3. An illustration of residual block.
	\end{figure}

	\noindent 
	\subsubsection{Dense-Cycle-GAN}
	
	When the source image is very different with target image, the major difficulty in modeling the transformation/translation/synthesis is that the location, structure, and shape of the source image and target image can vary significantly among different patients. In order to accurately predict each voxel in the anatomic regions, such as soft-tissue and bone structures in the task of mapping MRI to CT, inspired by densely connected CNN, several dense blocks are introduced to capture multi-scale information (including low-frequency and high-frequency) by extracting features from previous hidden layers and deeper hidden layers \cite{RN23}. As shown in generator architecture of Figure 4, the dense block is implemented by 5 convolution layers, a concatenation operator, and a convolutional layer to shorten the feature map size. Similar to the residual blocks used in Cycle-GAN, dense blocks are also often used as short skip connections in generators architecture.
	
	\begin{figure}
		\centering
		\noindent \includegraphics*[width=6.50in, height=4.20in, keepaspectratio=true]{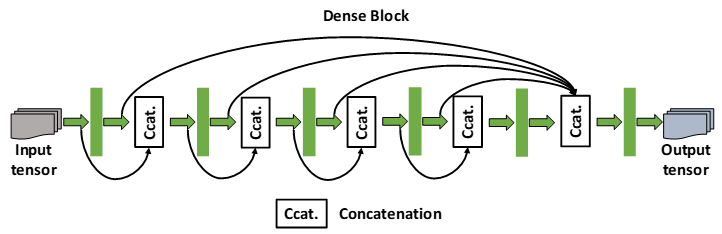}
		
		\noindent Fig. 4. An illustration of dense block.
	\end{figure}
	
	\noindent 
	\subsubsection{Unsupervised Image-to-Image Translation Networks (UNIT)}
	
	Some image synthesis problems such as multimodal MRI synthesis \cite{RN74} require multiple image domains mapping rather than mapping only one image domain to the other image domain. In computer vision, this kind of tasks were previously solved by unsupervised image-to-image translation networks (UNIT). UNIT aims at learning a joint distribution of images in different domains by using images from the marginal distributions in individual domains. It would be difficult to learn joint distribution from arrival marginal distribution without additional assumptions. To address the problem, a shared-latent space assumption was introduced and an unsupervised image-to-image translation framework based on coupled GANs were proposed \cite{RN20}.
	
	\noindent 
	\subsubsection{Bicycle-GAN}
	
	For multimodality image synthesis, Zhu et al. \cite{RN23} improved the UNIT by introducing Bicycle-GAN. This network aims to model a distribution of possible outputs in a conditional generative modeling setting. The ambiguity of the mapping is distilled in a low-dimensional latent vector, which can be randomly sampled at test time. A generator learns to map the given input, combined with this latent code, to the output. It was explicitly encouraged that the connection between output and the latent code to be invertible. This helps prevent a many-to-one mapping from the latent code to the output during training, also known as mode collapse, which produces more diverse results. Zhu et al. explored several variants of this approach by employing different training objectives, network architectures, and methods of injecting the latent code. Bicycle-GAN encouraged bijective consistency between the latent encoding and output modes. 
	
	\noindent 
	\subsubsection{StarGAN}
	
	StarGAN, also known as Unified GAN, is another variant of Cycle-GAN and used for multimodal image translation \cite{RN24}. The challenge of multimodal image synthesis is the limited scalability and robustness in handling more than two domains, since different models should be built independently for every pair of image domains. To address this limitation, StarGAN was proposed to solve the multiple modality image-to-image translations using only a single model. Such a unified model architecture of StarGAN allows simultaneous training of multiple datasets from different domains within a single network. This leads to superior quality of translated images compared to traditional Cycle-GAN models as well as the novel capability of flexibly translating an input image to any desired target domain. StarGAN was previously used in computer vision tasks such as a facial attribute transfer and a facial expression synthesis tasks \cite{RN24}, and was recently used for multimodal MRI synthesis \cite{RN74}.

	\noindent 
	\subsection{ Loss Function}
	
	As described above, the GANs rely on continuous improvement of generator network/networks and discriminator network/networks. The performance of these networks is directly dependent on the design of their loss functions.
	
	\noindent 
	\subsubsection{Discriminator Loss}
	
	Binary cross entropy (BCE) or signal cross entropy (SCE) losses are often used to supervise the discriminator network/networks \cite{RN2}. Since the goal of discriminator is to judge the authenticity of arrival synthetic image, the discriminator should improve its ability to discriminate/regard synthetic image as not real and the original image as real. For example, given source image $I_{src}$, the corresponding generator $F_G$ learned from previous iteration, the corresponding discriminator $F_D$, and its target original/ground truth image $I_{tar}$, the loss of discriminator measured by BCE can be expressed as follows:
	
	\begin{equation}
	F_D={arg\min}_{F_D}{\left\{BCE\left(F_D\left(I_{tar}\right),1\right)+BCE\left(F_D\left(F_G\left(I_{src}\right)\right),0\right)\right\}},
	\end{equation}
	where 1 denotes real and 0 denotes fake.
	
	\noindent 
	\subsubsection{Adversarial Loss}
	
	The loss function of generator is often composed of several losses for different purposes/constraints. Here we first discuss the adversarial loss. As introduced above, the goal of generator is to fool the discriminator, i.e., let discriminator think synthetic image is real, thus, given previously learned discriminator $F_D$, source image $I_{src}$, ground truth target image $I_{tar}$ and generator $F_G$, the adversarial loss measured by BCE can be expressed as follows:
	
	\begin{equation}
	L_{adv}=BCE\left(F_D\left(F_G\left(I_{src}\right)\right),1\right),
	\end{equation}
	which means by minimizing loss term of Eq. (2), the synthetic image $F_G(I_{src})$ would close to real for discriminator $F_D$.
	
	\noindent 
	\subsubsection{Image Distance Loss}
	
	There are several image distance losses measuring between synthetic image and target image. Recently, two kinds of widely used losses are pixel-wise loss and structural loss. MAE and mean square error (MSE) are often used as pixel-wise loss for image synthesis tasks \cite{RN38, RN75}. Some other works used $l_p$-norm ($p\in(1,2)$) as pixel-wise loss \cite{RN23}. As the $l_p$-norm regularization has fewer solutions than $l_2$-norm optimization (MSE), over-smoothing results (i.e. blur region in MSE loss optimization) are reduced. On the other hand, it is demonstrated that the optimization solution under $l_p$-norm regularization has more solutions than $l_1$-norm optimization (MAE). It means the misclassification situations (the solution on ±1) are minimized by averaging several solutions obtained by similar samples (the solution around ±1).
	
	The second component of the image distance loss function is the gradient difference loss (GDL), which measures the structural similarity between synthetic image and ground truth target image. Between any two images X and Y, the GDL is defined as:
	
	\begin{equation}
	GDL\left(X,Y\right)=\sum_{i,j,k}\left\{\begin{matrix}\left(\left|X_{i,j,k}-X_{i-1,j,k}\right|-\left|Y_{i,j,k}-Y_{i-1,j,k}\right|\right)^2\\+\left(\left|X_{i,j,k}-X_{i,j-1,k}\right|-\left|Y_{i,j,k}-Y_{i,j-1,k}\right|\right)^2\ \\+\left(\left|X_{i,j,k}-X_{i,j,k-1}\right|-\left|Y_{i,j,k}-Y_{i,j,k-1}\right|\right)^2\\\end{matrix}\right\},
	\end{equation}
	where i, j, and k represent pixels in x-, y-, and z-axis, respectively.
	
	\noindent 
	\subsubsection{Histogram Matching Loss}
	
	Recently, in order to force the synthetic image to reach a similar histogram distribution level as that of ground truth image,  Lei et al. proposed a histogram matching loss, also called MaxInfo loss \cite{RN76}. MaxInfo loss is a measure of mutual dependency between two probability distributions
	
	\begin{equation}
	MaxInfo\left(X,Y\right)=\sum_{i,j,k}{p\left(X_{i,j,k},Y_{i,j,k}\right)\log{\frac{p\left(X_{i,j,k},Y_{i,j,k}\right)}{p\left(X_{i,j,k}\right)\bullet p\left(Y_{i,j,k}\right)}}},
	\end{equation}
	where $p\left(X,Y\right)$ is a joint probability function of X and Y. $p\left(X\right)$ and $p\left(Y\right)$ are marginal probability functions of of X and Y.

	\noindent 
	\subsubsection{Perceptual Loss}
	
	The challenge of some synthesis tasks is that the structure/edge boundary would be blurred due to the residual anatomical mismatch between the training deformed source image and ground truth target image \cite{RN76}. If only the image distance loss is used (e.g., MAE and GDE), GAN-based methods could not produce sharp boundaries as it mixes mismatches between source and target images during training. Perceptual loss is often used to enhance the boundary contrast and sharpness. The main idea of perceptual supervision \cite{RN3} is that feeding forward networks (i.e., generator/generators) could generate high-confidence fooling image (i.e., synthetic image) by using a perceptual loss that measures the perceptual and semantic difference between synthetic image and ground truth image. 
	
	The perceptual loss is defined by feature difference on high-level feature maps. These high-level feature maps were extracted from both target image and synthetic target image, via a network named feature pyramid network (FPN) architecture. For example, for lung CT synthesis tasks, the FPN can be pre-trained using the dataset of thoracic CT images and paired lung contours obtained from 2017 AAPM Thoracic Auto-segmentation Challenge \cite{RN31, RN6}. FPN, denoted by $F_s$, extracted multi-level feature maps from the ground truth target image (X) and the synthetic image (Y), respectively, i.e., $f_X=\bigcup_{i=1}^{N}{F_s^i\left(X\right)}$ and $f_Y=\bigcup_{i=1}^{N}{F_s^i\left(Y\right)}$, where N is the number of pyramid levels. The perceptual loss is defined as the Euclidean distance between the two feature maps, and calculated as: 
	
	\begin{equation}
	L_p\left(f_X,f_Y\right)=\sum^N_{i=1}\frac{\omega_i}{{C_i}{H_i}{W_i}{D_i}}||F^i_s(X)-F^i_s(Y)||^2_2,
	\end{equation}
	where $C_i$ denotes the number of feature map channels at ith pyramid level. $H_i$, $W_i$ and $D_i$ denotes the height, width, and depth of that feature map. $\omega_i$ is a balancing parameter for feature level i. Since the semantic information of the feature map at higher pyramid levels would be coarse, the weight for that level’s perceptual loss should be enlarged, thus it is often set by $\omega_i=p^{i-1}$ with $p\in(1,2)$ \cite{RN76}.

	\noindent 
	\section{Implementation}
	
	Raw data from clinical databases are usually not suited for network training. It is important to perform data preprocessing such as cropping or zero-padding or patching based on a setup of network input dimension and size, image normalization and data augmentation prior to network training.
	
	\noindent 
	\subsection{Network input dimension and size}
	Based on different goals, 3D and 2D medical images are usually the datasets. Depending on the network design and graphics processing unit (GPU) memory limitation, some methods directly use the whole volume as input to train the network \cite{RN53}, while some methods process the 3D image slice by slice, called as 2.5D \cite{RN38}, rest works used 2D/3D patches \cite{RN97, RN9, RN76, RN103}. The 3D-based approaches take 3D patches or whole volume as input and utilize 3D convolution kernels to extract spatial and contextual information from the input images. Full-sized whole volume training often leads to increasing computational cost and complexity as larger number of layers are used. Compared to whole volume-based methods, some methods extract 2D small patches from 3D image by sliding 2D window across original images prior to network prediction, and then use patch fusion to obtain the final full-sized segmentation. 2D/3D patch-based methods are less computational demanding. 
	
	\noindent
	\subsection{Pre-processing}
	Pre-processing plays an important role in synthesis tasks, since there are intensity, contrast and noise variation in the images. To ease the network training, pre-processing techniques are usually applied prior to network training. Typical pre-processing techniques include registration \cite{RN100, RN81, RN26, RN78, RN1,RN98, RN12, RN15, RN96}, bias/scatter/attenuation correction \cite{RN49, RN50}, voxel intensity normalization \cite{RN243} and cropping \cite{RN44} etc.
	
	\noindent
	\subsection{Data augmentation}
	Data augmentation is used to reduce over-fitting and increase the amount of training samples. Typical data augmentation techniques include rotation, translation, scaling, flipping, distortion, linear warping, elastic deformation, and noise contamination \cite{RN38}.
	
	\noindent 
	\section{CGAN and Cycle-GAN Applications}
	Recently, cGAN and Cycle-GAN were successfully used for several medical/biomedical applications. In this subsection, we briefly summarize some recent cGAN and Cycle-GAN applications. These applications will also be discussed in detail in the next few chapters. The reviewed articles were categorized into two main groups in this study based on their study objectives: medical image synthesis tasks and biomedical image synthesis tasks. In each group, there are subgroups that specify the imaging modalities and clinical applications.

	\noindent
	\subsection{Medical Image Synthesis}
	\noindent
	\subsubsection{Multi-modality MRI Synthesis}
	MRI is widely used in clinical practice attribute to its capability in providing meaningful anatomical and functional information \cite{RN20, RN3825, RN3827, RN3837}. Through applying different MRI pulse sequences, multi-contrast images can be acquired while scanning the same anatomy. These images offer physicians complementary information for assessing, diagnosing, and planning treatment of various diseases. For instance, in brain MRI scans, T1-weighted (T1) images show distinguishable white and grey matters. T1-weighted and contrast-enhanced (T1c) images can be used for assessment of the change of tumor shape with its enhanced demarcation around tumor. T2-weighted (T2) images show fluid obviously from cortical tissue, while contours of lesion can be delineated clearly on fluid-attenuated inversion recovery (FLAIR) images \cite{RN1736, RN1712}. Therefore, integrating the strengths of each modality can help unveiling rich underlying information of tissue that facilitates diagnosis and treatment management \cite{RN1, RN23, RN32, RN55, RN2032, RN1596}. However, in MRI scan, due to limited scan time, inconsistent machine settings, scan artifacts and corruption, and patient allergies to contrast agents, it is difficult to apply a unified group of scan sequences to each individual patient even with a similar disease, for example, glioblastoma. The various imaging protocols for different patients across different institutes result in a lack of consistent image modalities for all patients, bringing challenges for clinical practice and longitudinal research. To tackle this challenge, cross-modal image synthesis has been proposed and widely investigated as an encouraging solution to generate the missing modalities taking in the available ones as input \cite{RN1721}.
	
	Recently, cGAN and Cycle-GAN frameworks have been investigated for multimodal MRI synthesis \cite{RN1721, RN1717, RN1852, RN1707, RN1734, RN2005, RN1728, RN1735}. These methods can be grouped into three main categories depending on their input/output modalities: 1) single-input single-output (SISO), 2) multi-input single-output (MISO), 3) multi-input multi-output (MIMO). 
	
	In SISO, a target image is generated from a given source image. Yu et al. \cite{RN1728} elevated the capability of the promising image-to-image translation algorithm cGAN for 3D MR image synthesis and  Flair image generation from T1 image. The competition between the generator and discriminator resulted in an achievable Nash equilibrium. Later, the same group proposed edge-aware GANs (Ea-GANs) to overcome the discontinuous synthesis across slices in 2D cGANs through capturing image context in a global level with a 3D estimation. In Ea-GANs, edge information was effectively preserved along with voxel-wise intensity to improve the synthesis performance (e.g., efficiency, image quality) by incorporating the edge information into the objective function of the generator, enforcing the generated image to have a similar edge map as the ground-truth or by integrating the edge information into both generator and discriminator. Very recently, they incorporated the sample-adaptive strategy into GAN models aiming to obtain the local sample space mapping for each individual input sample to improve the quality of synthesis \cite{RN1883}. Dar et al. \cite{RN1717} further studied the unimodal MR image synthesis between T1 and T2 using cGANs, where the models were trained with pixel-wise and perceptual losses in the case of given spatially registered image pairs and cycle loss for unregistered pairs, and tested  on brain MRI in both glioma patients and healthy subjects with improved synthesis accuracy. These SISO methods are generally optimized for capturing a unique correlation between the source and target modalities. 
	
	MISO approaches are proposed to overcome limitations of SISO when the source and target images are weakly correlated through learning the shared latent representations among multiple source images. Olut et al. \cite{RN1707} investigated a cGAN based approach for synthesizing MR angiography from T1 and T2 images. Joyce et al. \cite{RN1731} proposed an encoder-decoder network to synthesize Flair from multiple contrasts including T1, T2, and diffusion weighted imaging (DWI). A scalable GAN-based model was developed to flexibly take arbitrary subsets of the multiple modalities as input and generate the target modality \cite{RN1852}. 
	
	MIMO was firstly proposed by Chartsias et al. \cite{RN2068} using a deep fully convolutional neural network taking all the available MRI modalities as input and simultaneously synthesizing one or more missing modalities. A variety of GANs\cite{RN1734, RN1735} were also investigated for multimodal MR image synthesis from multi-contrast images. 
	
	\noindent
	\subsubsection{MRI-only Radiation Therapy Treatment Planning}
	MRI has superior soft tissue contrast over CT, allowing for improved organ-at-risk segmentation and target delineation for radiation therapy treatment planning \cite{RN33, RN25, RN4}. Since dose calculation algorithms rely on electron density maps generated from CT images for calculating dose, MRIs are typically registered to CT images and used alongside the CT image for treatment planning \cite{RN27}. Since electron density information and CT images are vital to the treatment-planning workflow, methods which generate electron density and CT image from MRIs, called sCT generation, have been investigated recently \cite{RN36, RN37, RN35}.
	
	CGAN has been used in the generation of sCT by introducing an additional discriminator to distinguish the sCT from real CT, improving the final sCT imaging qualities in comparison to the previously deep learning-based methods \cite{RN55}. GAN-based methods still require the training pairs of MRI and CT images to be perfectly registered, which can be difficult to carry out with the high levels of accuracy needed for image synthesis \cite{RN98}. If the registration has some local mismatch between the MRI and CT training data, i.e. soft tissue misalignment after bone-based rigid registration, cGAN-based methods would produce a degenerative network, decreasing their accuracy. Wolterink et al. show that training with pairs of spatially aligned MRI and CT images of the same patients is not necessary for Cycle-GAN-based sCT generation method \cite{RN98}.

	\noindent
	\subsubsection{CBCT Improvement/Enhancement}
	The incorporation of CBCT onto medical linear accelerators has allowed 3D daily image guidance for radiation therapy, enhancing the reproducibility of patient setup. CBCT is typically used either daily or weekly to verify patient setup and to monitor patient changes over the course of treatment. While CBCT is an invaluable tool for image guidance, the physical imaging characteristics, namely a large scatter-to-primary ratio, lead to image artifacts such as streaking, shading, cupping, and reduced image contrast. All of these factors prevent quantitative CBCT, hindering full utilization of the information provided by frequent imaging \cite{RN10, RN11}. The incorporation of CBCT into the clinic has led to increased interest in adaptive radiation therapy (ART), where dose would be calculated daily based on the patient’s true setup on the treatment table. ART could mitigate patient setup errors and account for day-to-day patient changes such as weight loss or inflammation. Removal of these uncertainties could allow for decreased margins on target volumes and increased sparing of organs at risk, potentially leading to higher target doses \cite{RN12}.
	
	Recently Cycle-GAN framework is used for CBCT correction due to its ability of efficiently converting images between the source domain and the target domain when the underlying structures are similar, even if the mapping between domains is nonlinear \cite{RN21}. Residual blocks were integrated into Cycle-GAN framework to enforce the learner to minimize a residual map between the CBCT (source) and the desired planning CT (target). Liang et al. also implemented a Cycle-GAN which achieved good performance at head-and-neck site \cite{RN51}.

	\noindent
	\subsubsection{Low-count PET and PET Attenuation Correction}
	Image synthesis among different PET images has been proposed to facilitate PET attenuation correction (AC) and low-count PET reconstruction. For the PET AC, cGANs and Cycle-GANs are used to directly estimate AC PET from non-AC PET (NAC PET). Dong et al. applied the Cycle-GAN to perform PET AC on whole body for the first time \cite{RN2399}. They also demonstrated the reliability of their method by including sequential scans in their testing datasets to evaluate the PET intensity changes with time on their AC PET as well as ground truth. 
	
	Low-count PET has extensive application in pediatric PET scan and radiation therapy response evaluation with advantage of better motion control and low patient dose. However, the low count statistics would result in increased image noise, reduced contrast-to-noise ratio, and large bias in uptake measurement. The reconstruction of a standard- or full-count PET from the low-count PET cannot be achieved by simple postprocessing operations such as denoising since lowering radiation dose changes the underlying biological and metabolic process, leading to not only noise but also local uptake values changes \cite{RN1837}. Moreover, even with a same tracer injection dose, the uptake distribution and signal level can vary greatly among patients. Cycle-GAN was applied as the learning-based low-count PET reconstruction methods \cite{RN8}, which was performed directly on low-count PET to generate full-count PET.
	
	Although Cycle-GAN demonstrates the feasibility of mapping low-count PET to full-count PET, a few studies investigated using both PET and MRI image as dual input channels to further improve the results when MR images are available. As expected, the addition of MRIs that provides anatomical information could help improve the performance of the network than without MRI. Chen et al. showed that their network was able to achieve 83\% accuracy when using only PET as input, and 89\% when using PET and MRI (PET+MR), in a clinical reading study of uptake status \cite{RN5107}. The potential reason of such difference lies in that the results by PET+MR were superior in reflecting the underlying anatomic patterns. The contribution of MR images was also validated in the study of Xiang et al by a significant improved PSNR \cite{RN4988}. They commented that structural information from MRIs yielded important cues for estimating the high-quality PET, even though structural MRIs differed from PETs significantly regarding their appearances.
	
	\noindent
	\subsection{Biomedical Image Synthesis}
	\noindent
	\subsubsection{Cell Synthesis}
	
	Cell image synthesis is often considered a three-step approach with the initial generation of cell phantoms, texture synthesis and a final simulation of the imaging system \cite{RN25}. The morphology can be modeled, e.g., using prior knowledge-based deformations of basic shapes \cite{RN25}, statistical shape models \cite{RN26}, spherical harmonics \cite{RN27} or using shape spaces derived from diffeomorphic measurements \cite{RN28}. As a next step, generated phantoms need to be translated to realistically looking images, which can be obtained either conventionally by mathematic description of the texture synthesis, by modeling protein distributions in sub-cellular components or by transfer of real textures to the simulated objects \cite{RN29}.
	
	With the advent of cGANs and Cycle-GAN, these networks demonstrate to excel at realistic image data generation as well. Extensions of the cGAN framework with conditional labels allow generating realistic images that reflect semantic properties provided to the generator \cite{RN20, RN23}. Recently, these methods were also used for generating biological images like multi-channel data of human cultured cells \cite{RN30}, protein localization in different cell cycle stages \cite{RN31} or entire tissues \cite{RN32}. In addition to generation of realistic textures, cGANs have also been successfully used to mimic the shape of cells in 3D \cite{RN33}. The simulations are usually finalized by placing synthetic phantoms in a virtual image space and by performing a simulation of the imaging system. This is often accomplished by adding artificial disruptions like dark current, photon shot noise, sensor readout noise and a point spread function \cite{RN34} or using more elaborate physically motivated wave-optical simulation approaches \cite{RN35}. 
	
	\noindent
	\section{Summary and Discussion}
	
	GANs have been increasingly used in the application of medical/biomedical imaging. As reviewed in this chapter, cGAN- and Cycle-GAN-based image synthesis is an emerging active research field with all these reviewed studies published within the last few years. With the development in both artificial intelligence and computing hardware, more GAN-based methods are expected to facilitate the clinical workflow with novel applications. Compared with conventional model-based methods, GAN-based methods are more generalized since the same network and architecture for a pair of image modalities can be applied to different pairs of image modalities with minimal adjustment. This allows easy extension of the applications using a similar methodology to a variety of imaging modalities for image synthesis. GAN-based methods generally outperform conventional methods in generating more realistic synthetic images with higher similarity to real images and better quantitative metrics. In implementation, depending on the hardware, training a GAN-based model usually takes several hours to days. However, once the model is trained, it can be applied to new patients to generate synthetic images within a few seconds or minutes. Due to these advantages, GAN-based methods have attracted great research and clinical interest in medical imaging and biomedical imaging.
	
	Although the reviewed literatures show the success of GAN-based image synthesis in various applications, there are still some open questions that need to be answered in future studies. Firstly, for the training of GAN-based model, most of the reviewed studies require paired datasets, i.e., the source image and target image need to have pixel-to-pixel correspondence. This requirement poses difficulties in collecting sufficient eligible datasets, as well as demands high accuracy in image registration. As compared to cGAN, it is demonstrated that Cycle-GAN can relax the requirement of the paired datasets to be unpaired datasets, which can be beneficial for clinical application in enrolling large number of patient datasets for training. However, even the image quality derived by Cycle-GAN can be better than cGAN, the numerical performance may not be improved significantly in some synthesis tasks due to the residual mismatch between synthetic image and ground truth target image.
	
	Secondly, although the merits of GAN-based methods have been demonstrated, its performance can be inconsistent under the circumstances that the input images are drastically different from its training datasets. As a matter of fact, unusual cases are generally excluded in most of the reviewed studies. Therefore, these unusual cases, which do happen occasionally in clinic setting, should be dealt with caution when using GAN-based methods to generate synthetic image. For example, some patients have hip prosthesis. The hip prosthesis creates severe artifacts on both CT and MR images. The related effect of its inclusion in training or testing dataset towards network performance is an important question that has not been studied yet. There are more unusual cases that could exist in all those imaging modalities and are worth of investigation, just to name a few: all kinds of implants that introduce artifacts, obese patients whose scan has higher noise level on image than average, and patients with anatomical abnormality. To conclude, the research in image synthesis is still wide open. The authors are expected to see more activities in this domain for the years to come.

	\noindent 
	\bigbreak
	{\bf Disclosures}
	
	The authors declare no conflicts of interest.

	\noindent 
	
	\bibliographystyle{plainnat}  
	\bibliography{arxiv}      
	
\end{document}